\documentclass[twocolumn,superscriptaddress,preprintnumbers,amsmath,amssymb,pre]{revtex4}

\usepackage{graphicx}
\usepackage{bm}

\begin{document}

\preprint{}

\title{Scaling theory of transport in complex networks}

\author{Lazaros K. Gallos}
\author{Chaoming Song}
\affiliation{Levich Institute and
Physics Department, City College of New York, New York, NY 10031, USA}
\author{Shlomo Havlin}
\affiliation{Minerva Center and Department of Physics, Bar-Ilan University, 52900 Ramat-Gan, Israel}
\author{Hern\'an A. Makse}
\affiliation{Levich Institute and
Physics Department, City College of New York, New York, NY 10031, USA}

\date{\today}

\begin{abstract}
Transport is an important function in many network systems and understanding
its behavior on biological, social, and technological networks is crucial
for a wide range of applications. However, it is a property that is not
well-understood in these systems and this is probably due to the lack of
a general theoretical framework. Here, based on the finding that
renormalization can be applied to bio-networks,
we develop a scaling theory of transport in self-similar networks.
We demonstrate the networks invariance under length scale renormalization
and we show that the problem of transport can be
characterized in terms of a set of critical exponents.
The scaling theory allows us to determine the influence of the modular
structure on transport.
We also generalize our theory by presenting
and verifying scaling arguments for the dependence of transport on microscopic features,
such as the degree of the nodes and the distance between them.
Using transport concepts such as diffusion and resistance we exploit this invariance
and we are able to explain, based on the topology of the network,
recent experimental results on the broad flow distribution in metabolic networks.
\end{abstract}

\maketitle

\section{Introduction}

Transport in complex networks is a problem of much interest in many
aspects of biology, sociology and other disciplines. For example, the study of metabolic fluxes in
organisms is crucial for a deeper understanding of how the cell carries
its metabolic cycle \cite{almaas}.
The use of metabolic flux analysis can provide important cellular physiological
characteristics using the network stoichiometry and predict optimal flux
distributions that satisfy a defined metabolic objective.
Similarly, information flow between the molecules of a
biological network provides insight for both the network structure
and the functions performed by the network. Such an example is the
concept of the `diffusion distance' in a protein-protein interaction network
which is used to predict possible interactions between proteins, simply
by studying diffusion in the existing network \cite{paccanaro}.
In food webs, energy
transfer between different levels of the web is crucial for the organism survival, while spreading
of a disease between different organisms may affect the regular operation
of the equilibrated system. Moreover, applications
of transport in complex networks extend to a plethora of other systems,
such as, for instance, transport of information in the Internet, spreading of diseases
and/or rumours in social networks, etc.
Despite its significance, the laws of transport in such a complex substrate
are yet unclear compared to transport in random media \cite{hba87,bah00}. This is due to the
complexity added by the heterogeneous degree distribution in such networks.


We study transport in real-world biological networks and via a model,
which possess both self-similar properties and the scale-free
character in their degree distribution. We explain our
results with theoretical arguments and simulation analysis.
We use approaches from renormalization theory in statistical physics that enable us to exploit the
self-similar characteristics of the fractal networks and develop a scaling theory of transport,
which we use to address the effects of the modularity and the
degree inhomogeneity of the substrate.

Due to the existence of a broad degree distribution, transport on a network is different
when it is between two hubs with a large number of connections $k$  or between low-degree nodes.
We therefore characterize the transport coefficients by their explicit dependence on $k_1$,
$k_2$, and $\ell$, where $k_1$ and $k_2$ denote 
the degree of two nodes ($k_1>k_2$), separated by a distance $\ell$ (distance is measured
by the minimum number of links, i.e. it is the chemical distance).
We study the diffusion time $T(\ell;k_1,k_2)$ and the resistance $R(\ell;k_1,k_2)$ between
any two nodes in the system. The dependence on $k_1$ and $k_2$ is not significant in homogeneous systems,
but is important in networks where the node degree spans a wide range of values,
such as in biological complex networks.
In fact, this dependence is critical for many other properties as has been already
shown for e.g. fractality, where traditional methods of measuring the fractal
dimension may fail because they do not take into account this inhomogeneity \cite{shm,shm2}.

Modularity is one of the most important aspect of these networks with direct implications
to transport properties. Here, we quantify the modular character of complex networks
according to our box-covering algorithm and reveal a connection between modularity
and flow.
Our results are consistent with recent experiments and metabolic flux studies,
and provide a theoretical framework to analyze transport in a wide variety of network systems.

\subsection{Metabolism modeling}

In metabolism modeling, there exist three main approaches \cite{Stelling}. (i) The most detailed analysis includes dynamic {\it mechanism-based} models \cite{Kitano},
but in general it is very difficult to incorporate experimental values for the needed kinetic parameters.
(ii) In the second approach one simplifies the above models, and calculates the fluxes in a metabolic network via flux
balance analysis, which includes a family of static {\it constraint-based} models \cite{Reed}. The limiting factor in this analysis is that the problem
is underconstrained (the number of unknown parameters, i.e. the fluxes, is larger than the number of
metabolite conservation equations) and cannot be solved uniquely. 
(iii) Finally, a third approach that is widely used in metabolism modeling is to ignore stoichiometry,
and focus only on the metabolites interactions without any thermodynamic aspects, which leads to {\it interaction-based} models \cite{Jeong},
i.e. undirected networks where a link connects two nodes that participate in a metabolic reaction.
In this paper we follow this third approach and we use this interaction-based network to
study transport on this network,
by drawing an analogy between the metabolic network and a resistance network. If we represent the
metabolites as nodes that are linked through electrical resistances and the current flow
represents the flux we can solve this problem without additional constraints, and
this solution may shed additional light on the involved processes. The advantage of our
approach is that it can isolate the topological effect
and we can address a
broader aspect of transport in biological networks, such as whether the observed flux inhomogeneity
is because of the network topology itself or due to the adopted flux constraints.
Moreover, this approach enables us to carry similar studies for diffusion on such networks.

\section{Modularity, diffusion and resistance}

\begin{figure}
\centering{ { (a) \resizebox{8cm}{!} {\includegraphics{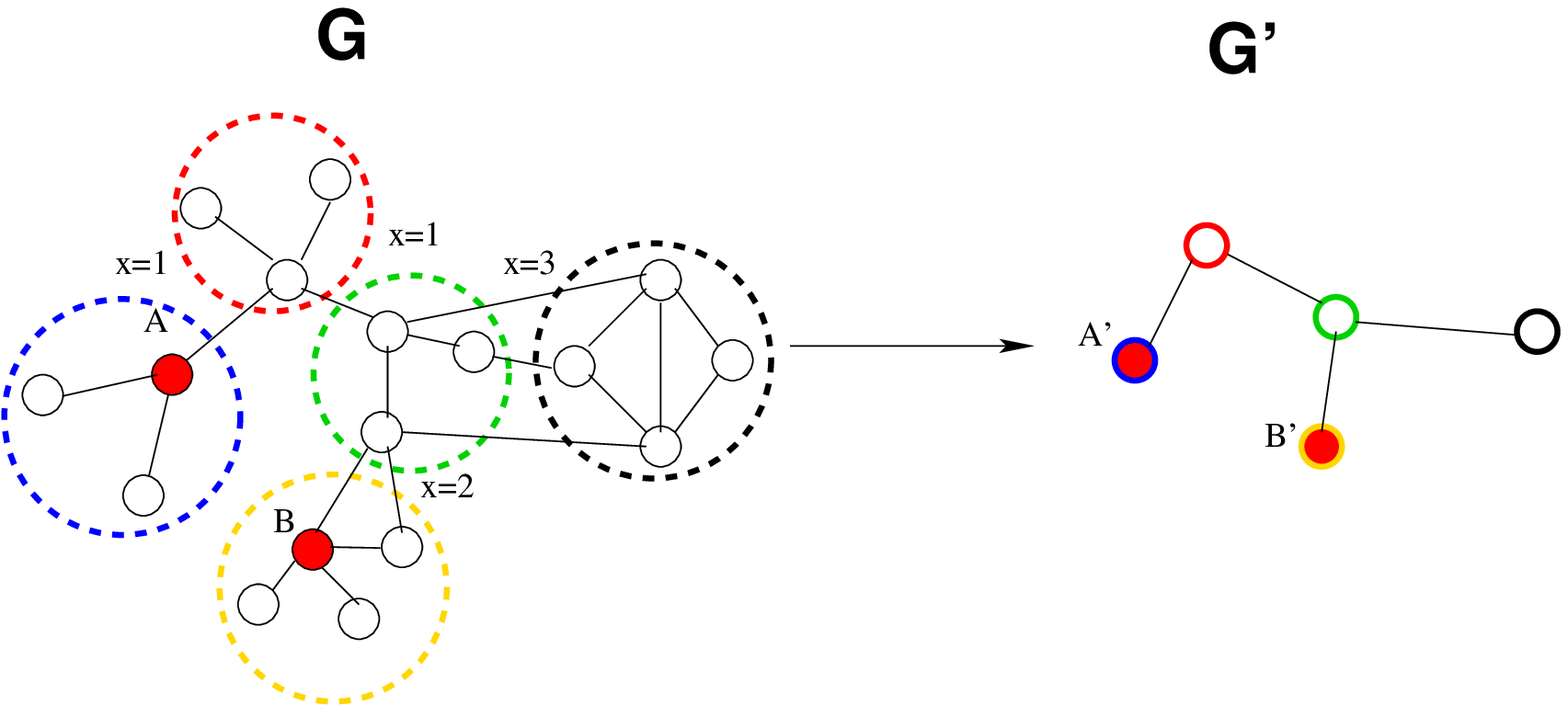}}}
{\resizebox{4.2cm}{!} {\includegraphics{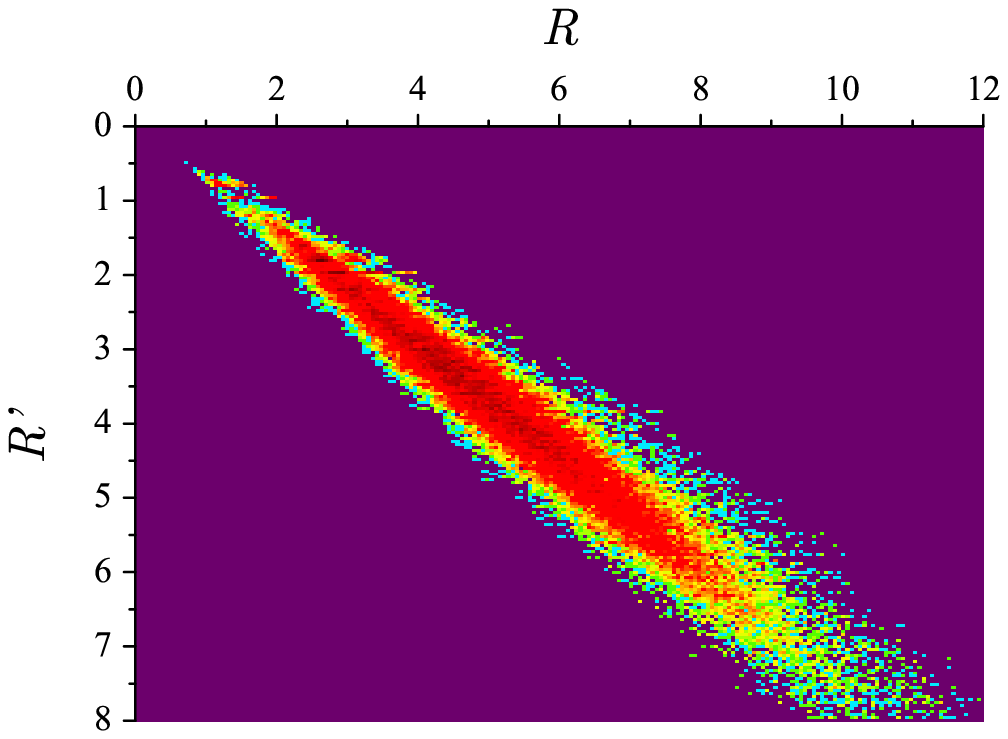}}}
{\resizebox{4.2cm}{!} {\includegraphics{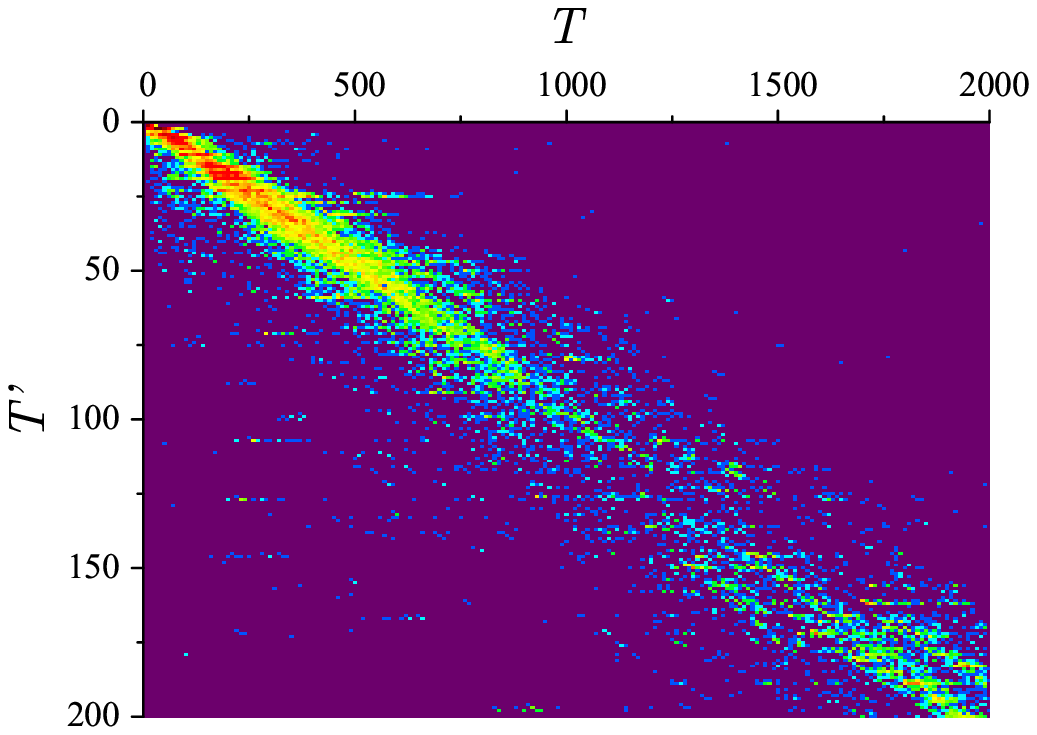}}}
(b) (c)\\ { \resizebox{8cm}{!} {\includegraphics{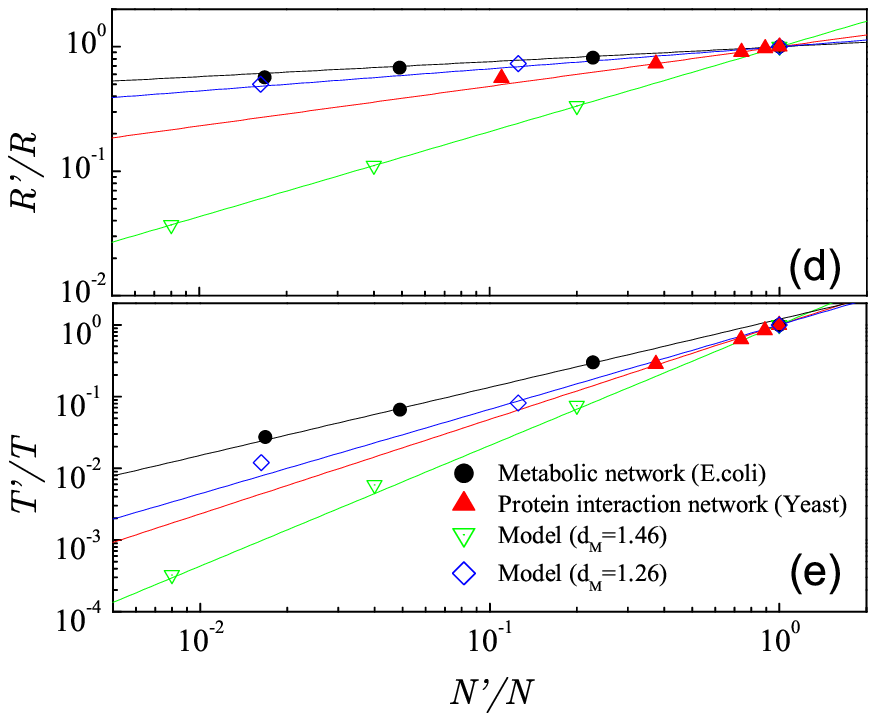}}}
}
\caption{\label{FIGrenor} (a) Example of a network $G$ that undergoes renormalization to a
network $G'$. For this example the maximum possible distance between any two nodes
in the same box has to be less than $\ell_B=3$. We compare the resistance $R(A,B)$ and diffusion time $T(A,B)$ between the two local-hub nodes A and B in $G$ to
the resistance $R'(A',B')$ and $T'(A',B')$ of the renormalized network $G'$. The values of $R'$ and $T'$ are
found to always be proportional to $R$ and $T$, respectively, for all pairs of nodes.
(b),(c) Typical behavior of the probability distributions for the resistance $R$ vs $R'$ and the diffusion time $T$ vs $T'$, respectively,
for a given $\ell_B$ value. Similar plots for other $\ell_B$ values verify
that the ratios of these quantities during a renormalization stage are roughly constant for all pairs of nodes in a given biological network.
(d),(e) We present the average value of this ratio for the resistance $R/R'$ and the diffusion time $T/T'$, respectively,
as measured for different $\ell_B$ values (each point corresponds to a different value of $\ell_B$).
Results are presented for both biological networks, and two fractal network models with different $d_M$ values.
The slopes of the curves correspond to the exponents $\zeta/d_B$ in (d) and $d_w/d_B$ in (e) (see Table 1).
}
\end{figure}

In our work, we focus on two different examples of biological networks, namely the {\it E.coli} metabolic
network \cite{Jeong} and the yeast protein interaction network (PIN) \cite{Han}.
Both networks have been shown to have fractal properties and
can be covered with $N_B(\ell_B)$ non-overlapping boxes,
where in each box the maximum distance between any two nodes is less than $\ell_B$,
the maximum distance in a box \cite{shm,shm2}. For a fractal network of $N$ nodes, $N_B$ follows a power-law dependence on $\ell_B$,
\begin{equation}
\label{EQnb}
N_B(\ell_B)/N \sim \ell_B^{-d_B}
\end{equation}
and defines $d_B$ as the fractal (or box) dimension of a network. These networks
are also self-similar, i.e. their main properties, such as the degree distribution,
remain invariant under a renormalization scheme where
each box is replaced by a (super) node and links between boxes are transferred
to the nodes of the renormalized network (see e.g. the example in Fig.~\ref{FIGrenor}a
for a network $G$, tiled with $\ell_B=3$ boxes, that yields the network $G'$).
Many biological networks in the
intermediate renormalized stages were shown to have similar properties as the original network.

This renormalization procedure also implies the presence of self-similar modularity in all
length-scales, which is a central feature of these networks. The term modularity refers to the existence of
sets of nodes whose links are connected preferably within this set rather than to the rest of the network.
Thus, after tiling a network for a given value of $\ell_B$, we introduce a measure of modularity for
a network as
\begin{equation}
M(\ell_B) = \frac{1}{N_B} \sum_{i=1}^{N_B} \frac{L_i^{\rm in}}{L_i^{\rm out}} \,,
\end{equation}
where $L_i^{\rm in}$ and $L_i^{\rm out}$ represent the number of links that start in a given
box $i$ and end either within or outside $i$, respectively. Large values of $M$ correspond,
thus, to a higher degree of modularity. Since the numerical value of $M(\ell_B)$ varies, though,
with $\ell_B$, a more reliable measure is the modularity fractal exponent $d_M$ which we define through:
\begin{equation}
M(\ell_B) \sim \ell_B^{d_M} \,.
\end{equation}
The value of $d_M=1$ represents the borderline case that separates modular ($d_M>1$)
from random non-modular ($d_M<1$) networks. For a lattice structure, the value of $d_M$ is 
exactly equal to $d_M=1$.

In Ref.~\cite{shm2} we had introduced a fractal network model where a network grows by adding
$m$ new offspring nodes to each existing network node, resulting in well-defined modules.
In that version of the model, modules are connected to each other through $x=1$ links,
which leads to a tree structure. A generalization
of this model (presented in detail in the Supporting Information) allows us to tune the degree of modularity in
the network by assigning a larger number of links $x>1$ between modules. While for $x=1$ all the modules
are well-defined, increasing $x$ leads to the presence of loops and to
a progressive merging of modules, so that for large $x$ values
a node cannot be assigned unambiguously in a given module and modularity is destroyed.
A straightforward analytical calculation in this case leads to (see Supporting Information)
\begin{equation}
\label{EQdef_dM}
d_M = \frac{\ln \left( 2 \frac{m}{x}+1 \right)}{\ln 3} \,.
\end{equation}
In this paper, we use this value of $d_M$ for the model and calculate
$d_M$ for real networks in order to study the influence of modularity on network transport.

In general, the problem of transport is expressed in terms of $T(\ell;k_1,k_2)$ and $R(\ell;k_1,k_2)$, where $T$ is the average
first passage time
needed by a random walker to cover the distance $\ell$ between two nodes with degrees $k_1$ and $k_2$, respectively, and $R$ represents the resistance between these two nodes.
For homogeneous systems (with very narrow degree distribution $P(k)$) such as lattices and regular fractals, there is no dependence on $k_1$ and $k_2$ and the average is only over the
distance $\ell$. One of the goals of this paper is to find the scaling of $T$ and $R$ in heterogeneous networks with a broad degree distribution and self-similar properties.

In the general case of a renormalizable network, $T$ and $R$ scale in the
renormalized network $G'$ (the primes always denote a quantity for the
renormalized network) as 
\begin{equation}
\label{EQratios}
T'/T = \ell_B^{-d_w} \,,\, R'/R = \ell_B^{-\zeta} \,.
\end{equation}
The exponents $d_w$ and $\zeta$ are the random walk exponent and the resistance exponent, respectively. This equation is valid as an average of $T$ and $R$ over the entire system,
applying for example in different generations when growing or renormalizing a fractal object.
Thus, this relation holds true for both homogeneous and inhomogeneous systems.

For homogeneous systems, the above exponents $d_w$ and $\zeta$ are related through the Einstein relation
\cite{hba87}
\begin{equation}
\label{EQeinstein}
d_w = \zeta + d_B \,,
\end{equation}
where $d_B$ is the fractal dimension of the substrate on which diffusion takes place.
This relation is a result of the fluctuation--dissipation theorem relating spontaneous
fluctuations (diffusion) with transport (resistivity) and the underlying
structure (dimensionality) \cite{hba87}. Although the validity of this relation
for scale-free networks is not yet clear, our following analysis shows that
it also applies for these systems, as well.


\section{Renormalization and scaling theory}

The renormalization procedure on self-similar biological networks provides a yet unexplored method for
estimation of the dynamical exponents in these systems.
Since such a network is left invariant after substituting all nodes in a box with a single node,
we can calculate the transport properties on the networks during successive renormalization stages.
With this method we can also study transport in biological networks before ($R$ and $T$) and
after renormalization ($R'$ and $T'$).
The results are presented in Fig.~\ref{FIGrenor} for the yeast PIN, the E.coli metabolic network and the
fractal network model with $d_M=1.46$ (tree, highly modular) and $d_M=1.26$ (network
with loops and lower modularity).
Figures \ref{FIGrenor}(b) and \ref{FIGrenor}(c) suggest a linear relation between $R'$ and $R$ ($T'$ and $T$)
for a given value of $\ell_B$, so that the ratio $R'/R$ (and $T'/T$) is almost constant for all boxes in the system
for this $\ell_B$ value.  For a different value of the box diameter $\ell_B$ this ratio is again constant
for all boxes in the network, but assumes a different value.
We can plot the values of this ratio as a function of the network
size ratio $N'/N$ for different values of $\ell_B$ (Figs.~\ref{FIGrenor}d and ~\ref{FIGrenor}e).
The data indicate the existence of a power-law dependence, and a comparison with the model
networks shows that the results are consistent with PIN exhibiting a more modular structure
compared to the metabolic network.

Although in principle we can use our numerical results to directly calculate the exponents $d_w$ and $\zeta$ through Eq.~\ref{EQratios}, this method is not practical because the variation of $\ell_B$ is very small. We can overcome this difficulty by using the system size $N$ instead, where we
combine Eqs.~(\ref{EQnb}) and (\ref{EQratios}) to get
\begin{equation}
\frac{T'}{T} = \left( \frac{N'}{N} \right)^{d_w/d_B} \,,\, \frac{R'}{R} = \left( \frac{N'}{N} \right)^{\zeta/d_B} \,.
\end{equation}
Thus, the slopes in Figs.~\ref{FIGrenor}d and \ref{FIGrenor}e correspond to the exponent ratios $d_w/d_B$ and $\zeta/d_B$, respectively.

Notice also that the verification of the above equation through Fig.~\ref{FIGrenor} validates the relation in
Eq.~(\ref{EQratios}) for inhomogeneous systems.
The numerical values for the calculated exponents are shown in Table~\ref{table1}.
These ratios are consistent in all cases, within statistical error, with the Einstein relation, Eq.~(\ref{EQeinstein}).

\begin{table}
\caption{\label{table1}Values of the exponents calculated from Fig.~\protect\ref{FIGrenor}}
\begin{tabular}{cccc}
Network & $d_B$ & $\zeta/d_B$ & $d_w/d_B$ \\
\hline
Metabolic network (E.coli) & 3.3 & 0.08 $\pm$ 0.1 & 0.98 $\pm$ 0.1 \\
PIN (yeast) & 2.2 & 0.3 $\pm$ 0.04 & 1.3 $\pm$ 0.04 \\
Model ($d_M=1.46$, $m/x=2/1$) & 1.46 & $\frac{\ln3}{\ln5}$ & $1+\frac{\ln3}{\ln5}$ \\
Model ($d_M=1.26$, $m/x=3/2$) & 1.89 & $\frac{1}{3} \left( \frac{\ln3}{\ln2}-1 \right)$ &
$\frac{1}{3} \left(\frac{\ln3}{\ln2}+2 \right)$ \\
\end{tabular}
\end{table}

Using these scaling arguments and the renormalization property
of these networks we next predict the dependence of both $R$ and $T$ on the
distance $\ell$ between two nodes and their corresponding degrees $k_1$ and $k_2$.
After renormalization the network becomes smaller, so that both the degrees and the distances in the
network decrease. A distance $\ell$ in $G$ is scaled by a factor $\ell_B$ in $G'$ so that
$\ell'=\ell/\ell_B$,
while in earlier work \cite{shm} it has been shown that the degree $k$ of the largest hub in a
box transforms to a degree $k'$ for the renormalized box, where
$k' = \ell_B^{-d_k} k$, and $d_k$ is an exponent describing the scaling of the degree.
According to the result of Fig.~\ref{FIGrenor} and Eq.~(\ref{EQratios}) it follows,
\begin{eqnarray}
\label{EQrrr}
R'(\ell' ; k_1',k_2') = \ell_B^{-\zeta} R(\ell ; k_1,k_2) \\
T'(\ell' ; k_1',k_2') = \ell_B^{-d_w} T(\ell ; k_1,k_2) \,.
\end{eqnarray}
Using dimensional analysis (see Supporting Information) we can show that
\begin{equation}
\label{EQfinalscaling}
R(\ell ; k_1,k_2) = k_2^{\zeta/d_k} f_R\left( \frac{\ell}{k_2^{1/d_k}} , \frac{k_1}{k_2}\right)
\end{equation}
\begin{equation}
\label{EQfinalTscaling}
T(\ell ; k_1,k_2) = k_2^{d_w/d_k} f_T\left( \frac{\ell}{k_2^{1/d_k}} , \frac{k_1}{k_2}\right) \,,
\end{equation}
where $f_R()$ and $f_T()$ are undetermined functions. In the case of homogeneous networks
where there is almost no $k$-dependence, these functions reduce to the forms $f_R(x,1)=x^\zeta$, $f_T(x,1)=x^{d_w}$, leading to the classical relations $R\sim\ell^\zeta$ and $T\sim\ell^{d_w}$.

\begin{figure}
\centering{ { \resizebox{8.8cm}{!} {\includegraphics{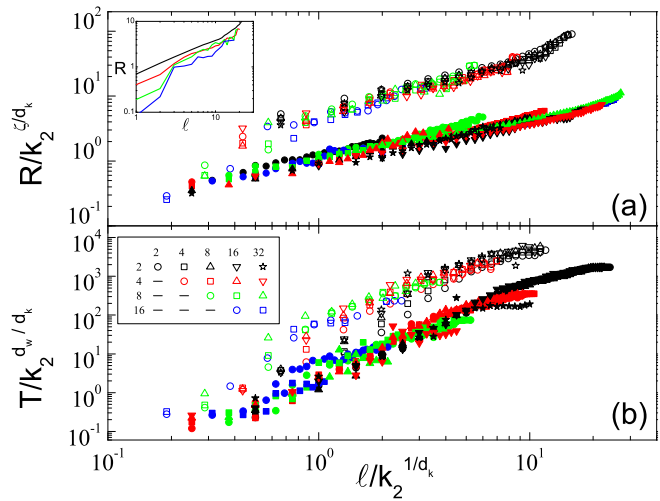}}}
}
\caption{\label{FIG_rw_res_rescaled} Rescaling of (a) the resistance and (b) the diffusion time
according to Eqs.~(\protect\ref{EQfinalscaling}) and (\protect\ref{EQfinalTscaling})
for the protein interaction network of yeast
(upper symbols) and the fractal network generation model (lower filled symbols).
The data for PIN have been vertically shifted upwards by one decade for clarity.
Different symbols correspond to different ratios $k_1/k_2$ and different colors denote
a different value for $k_1$. Inset: Resistance $R$ as a function of distance $\ell$,
before rescaling, for constant ratio $k_1/k_2=1$ and different $k_1$ values.
}
\end{figure}

The scaling in Eqs.~(\ref{EQfinalscaling}) and (\ref{EQfinalTscaling}) is supported by the numerical data collapse shown
in Fig.~\ref{FIG_rw_res_rescaled}. For the data collapse we used the values of the exponents $\zeta$ and $d_w$ as
obtained from the renormalization method above (Table \ref{table1}) confirming the scaling in Eqs.~(\ref{EQfinalscaling}) and (\ref{EQfinalTscaling}).

The functions $f_R$ and $f_T$ introduced in Eqs.~(\ref{EQfinalscaling}) and (\ref{EQfinalTscaling})
have two arguments,
so we first need to fix the ratio $k_1/k_2$ and in the plot (Fig.~\ref{FIG_rw_res_rescaled}) we
present different ratios using different symbols. We observe that the differences
among varying ratios are small, so that
the $k_1/k_2$ dependence in Eqs.~(\ref{EQfinalscaling}) and (\ref{EQfinalTscaling}) can be
neglected,
\begin{equation}
R(\ell ; k_1,k_2) = k_2^{\zeta/d_k} f_R\left( \frac{\ell}{k_2^{1/d_k}}\right) \,,
\end{equation}
with an analogous form for $T(\ell ; k_1,k_2)$. We are, thus, led to a simpler approximate form
where the values of diffusion and resistance between two nodes depend only on the lowest degree node
$k_2$ and the nodes distance $\ell$. This is a generalization to real networks of the result in Ref.~\cite{lopez}
for random model networks where, without taking into account the distance, the resistance was found to depend
solely on $k_2$.

\section{Influence of modularity on transport}

Modularity is a central feature of biological networks which contributes to
a more efficient use of resources in the network, with yet unclear consequences
for transport in these systems.
In this section we use the fractal network model, which reproduces the main
features of real networks, to better understand the influence
of modularity on transport.

A direct calculation of $\zeta$ for the fractal network model is as follows.
We consider the growth model where the distance $\ell$ between two nodes
increases by a factor of 3, i.e. $\ell'=3\ell$,
and the resistance $R$ between two neighbor nodes in $G$ increases by a
factor $3/x$ in $G'$, since the linear distance between the two nodes has increased
by a factor of 3 and there are $x$ parallel paths connecting these nodes,
i.e. $R'=3R/x$. Combining these equations with Eq.~(\ref{EQratios}) we
find that the exponent $\zeta$ for this model is given by:
\begin{equation}
\label{EQzetadx}
\zeta = \frac{\ln (3/x)}{\ln 3} = 1 - \frac{\ln x}{\ln 3} \,.
\end{equation}
Notice that for a tree structure ($x=1$) we get $\zeta=1$ as expected.
This result is also an important step in linking
statics with dynamics (a long standing problem in percolation theory \cite{hba87,bah00}).
Using only the value of $x$ that describes how self-similar modules are connected to each other
we can directly obtain the dynamic exponent $\zeta$,
i.e. how the structural property of modularity affects dynamics.

For the fractal dimension of model networks we already know that \cite{shm2}
\begin{equation}
d_B = \frac{\ln(2m+x)}{\ln 3} \,.
\end{equation}
If we assume that the Einstein relation in Eq.~(\ref{EQeinstein}) is valid, then we can
also calculate the value for the random walk exponent $d_w$:
\begin{equation}
\label{EQfinaldw}
d_w = \frac{\ln\left( \frac{6m}{x}+3\right)}{\ln 3} \,.
\end{equation}
A comparison of Eq.~(\ref{EQfinaldw}) with Eq.~(\ref{EQdef_dM}) yields:
\begin{equation}
\label{EQdw_dM}
d_w = 1 + d_M \,.
\end{equation}
This is a very simple, yet powerful result. It manifests that for the fractal network model
the degree of modularity directly affects the efficiency of transport and is the
main feature that controls the type of diffusion.

\begin{figure}
\centerline{ \resizebox{8cm}{!} {\includegraphics{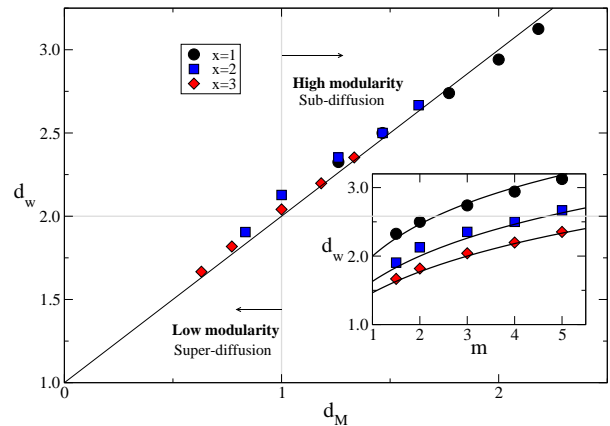}} }
\caption{\label{FIG_dw_m}
Comparison of the random walk exponent $d_w$ extracted numerically (symbols) with the theoretical prediction (Eq.~(\protect\ref{EQdw_dM}), line) vs the modularity
exponent $d_M$, for different values of $m$ and $x$.
Inset: Direct (unscaled) numerical calculation of $d_w$ as a function of $m$, for varying $x$ values (shown in the plot).
}
\end{figure}

The above relation, Eq.~(\ref{EQdw_dM}), is verified in Fig.~\ref{FIG_dw_m}
for the fractal network
model. We generate a number of model networks where 
we vary both the number of loop-forming links $x$ and the number of offsprings $m$.
For each pair of $m$ and $x$ we calculate numerically the exponent $d_w$ from the slope
of figures similar to Figs.~\ref{FIGrenor}d and \ref{FIGrenor}e and use Eq.~(\ref{EQdef_dM})
for the value of $d_M$. The results are fully consistent with Eq.~(\ref{EQdw_dM})
and all the points lie on the predicted line.
Sub-diffusion ($d_w>2$) is observed for $d_M>1$, in accordance with our observation that
modularity slows down diffusion. On the contrary, for non-modular networks diffusion
is accelerated remarkably ($d_w<2$) which is also in agreeement with previous work
on random networks. When $d_M=1$ we recover classical diffusion ($d_w=2$), even though
the structure is still that of a scale-free network.

\section{Flow distribution across the network}

In our scaling theory above we derived results for the average values of the current flowing in
a complex network.
The inherent inhomogeneity and modularity of biological networks is expected, though, to strongly
influence the distribution of flow throughout the network.
Using flux balance analysis, it was recently shown that the distribution
of fluxes in the metabolic network is highly uneven and a small number
of reactions have the largest contribution to the overall metabolic flux activity \cite{almaas}.
To study the influence of the complex substrate on the flow distribution
we calculate the probability $P(I)$ of current $I_{ij}$ flowing between all two neighboring nodes
i and j.

\begin{figure}
\includegraphics{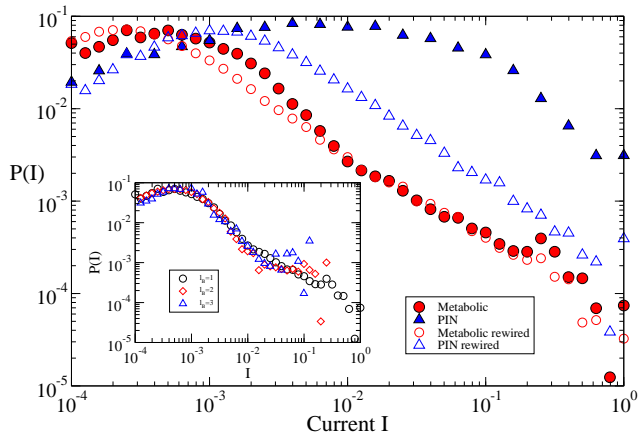}
\caption{\label{FIG_Ipdf} Probability distribution $P(I)$ of current magnitudes $I$
flowing through the links in PIN (solid triangles) and metabolic networks (solid circles). Empty
symbols are the corresponding results for the randomly rewired networks.
Inset: Invariance of $P(I)$ for the metabolic network under renormalization with different
$\ell_B$ values.
}
\end{figure}

The probability distribution $P(I)$ for the magnitude of the current $I$ across a link
of the metabolic network decays according to a power law (see Fig.~\ref{FIG_Ipdf}). The form of the curve and
the exponent in the range 1.0-1.5 of the decay is very similar to those found in previous studies of the metabolic flux,
both experimental \cite{Emmerling} and theoretical \cite{almaas}.
The decay suggests that only a small fraction of link carries high current.
For the yeast PIN the distribution is even broader, and its form is different from
the metabolic network.
The variation of $P(I)$
is smaller in PIN compared to the metabolic network, meaning that in PIN
there is a larger number of important links that carry large currents.
The self-similar character of the biological networks is also verified for the distribution $P(I)$,
as well, which remains invariant after renormalizing the network, as seen in the inset of Fig.~\ref{FIG_Ipdf}.

Next we explore the connection between $P(I)$ and topology.
The information contained in $P(I)$ can be better understood if we compare these
results with a surrogate random case. The random case is obtained by rewiring the original networks,
preserving the degree distribution $P(k)$, but destroying any correlations between neighboring nodes.
Thus, we remove all traces of the initial network organization. The distribution $P(I)$ for the metabolic network
remains almost the same under this rewiring, indicating that, despite the modular character of the network,
the original structure behaves similar to uncorrelated networks, since the degree correlations do not
affect $P(I)$ in the metabolic network. In contrast, the
distribution $P(I)$ for the rewired PIN is very different than the original
distribution and is similar to that of the metabolic network. We can, thus, conclude
that the original PIN has a much richer structure that deviates from the random case,
corroborating thus the results on modularity from the previous section on diffusion.

\begin{figure}
\includegraphics{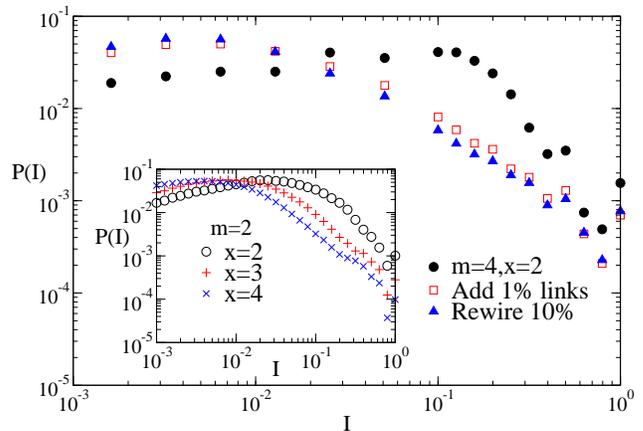}
\caption{\label{FIG_model_Ipdf} Probability distribution $P(I)$ for the fractal model
before and after randomly adding 1\% of links or rewiring 10\% of the network.
Inset: $P(I)$ for the fractal model with varying $d_M$ values,
where $m=2$ and $x$ varies from 2 to 4.
}
\end{figure}

The above results for the biological networks can be understood in terms of
the fractal network model, where we can control the number of links that form
loop structures in a network.
In Fig.~\ref{FIG_model_Ipdf} we calculate $P(I)$ for the model network with
$m=4$ and $x=2$, which is a highly modular structure ($d_M=1.46$).
The form of the $P(I)$ distribution is similar to that of the PIN, but if we add
a small number of random links (or equivalently rewire a small part of the network)
this distribution is significantly influenced in a similar way as observed
in random PIN rewiring. This suggests again that in the case of PIN modularity is high.
In the inset of this Figure
we can also see that as $x$ increases, i.e. more loops appear
in the structure, the distribution has a longer tail, which shows that there is a smaller
number of high-current links.

Since the number of added links in Fig.~\ref{FIG_model_Ipdf} is small, the
modularity is preserved. We verified that the main factor that influences
$P(I)$ is the number of loops in the network, rather than modularity itself,
by fixing $d_M$ and only varying $x$.
In this case (described in Supporting Information) the $P(I)$ distribution
is different for networks with the same $d_M$ exponent.
This can also be seen through
Eq.~(\ref{EQzetadx}), where the resistance exponent depends only on $x$.

\begin{figure}
\centering{
{\resizebox{7cm}{!} {\includegraphics{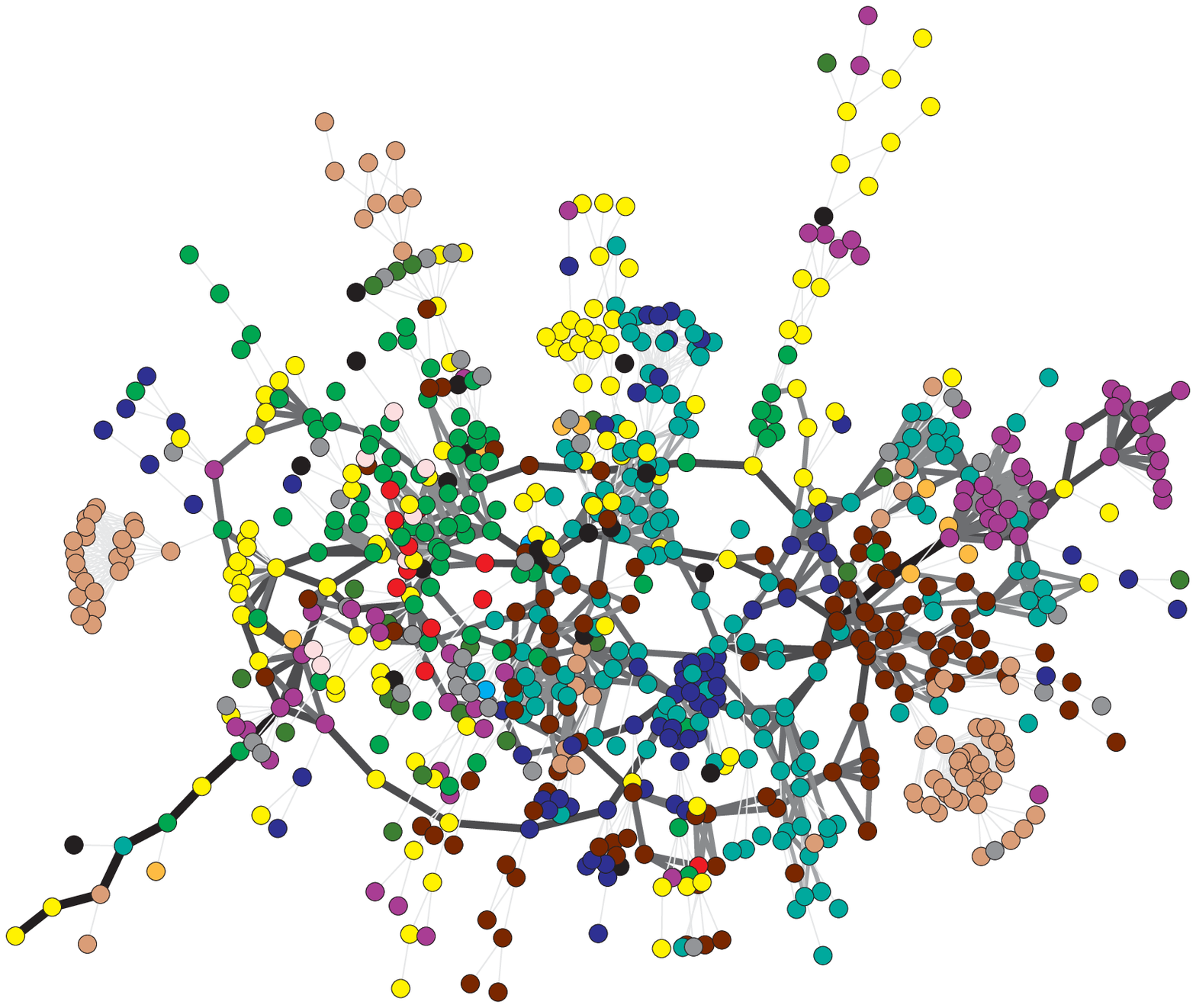}}}
{\resizebox{6cm}{!} {\includegraphics{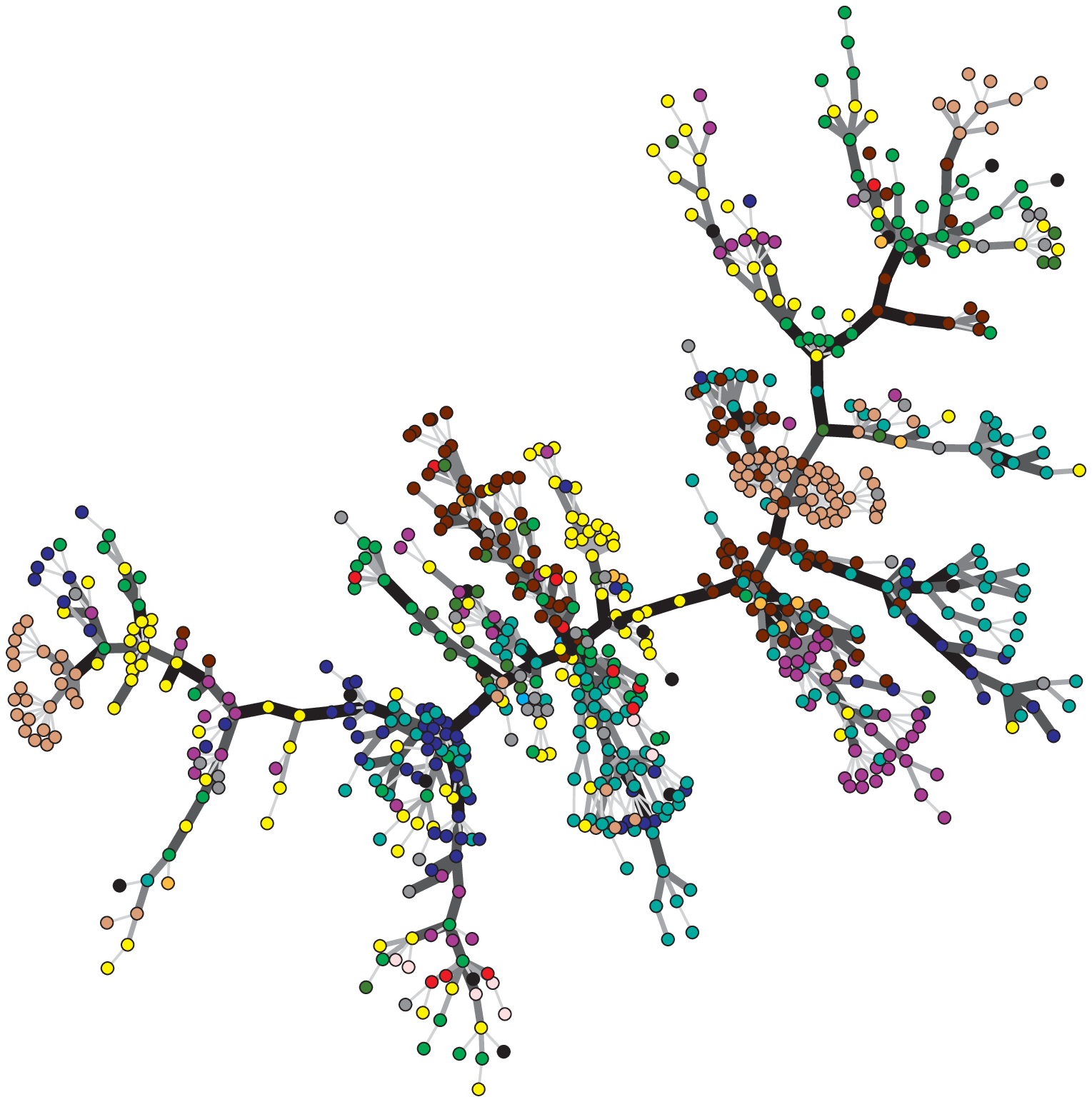}}}
{\resizebox{2cm}{!} {\includegraphics{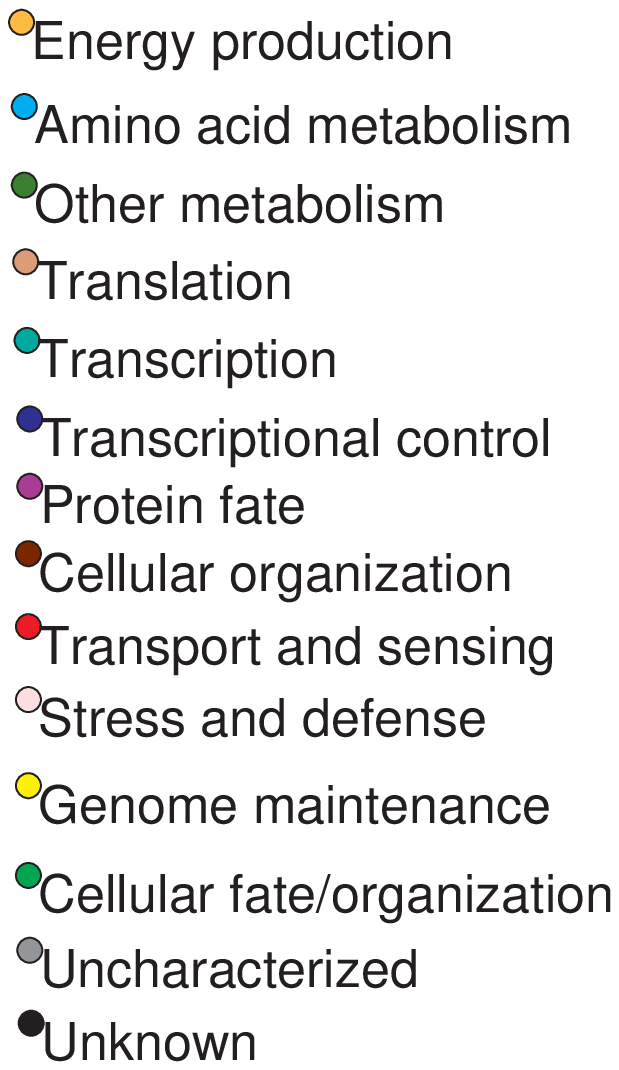}}}
}
\caption{\label{FIG_mst} (a) Current flow through the links of the yeast PIN network, for one random
selection of the two nodes acting as current input/output. (b) Minimum spanning tree for the PIN.
The thickness of a link corresponds to the current flowing through this link.
Different node colors correspond to different protein functions. 
}
\end{figure}

Using the information of $I_{ij}$ we can also construct the `backbone'
of the network in the form of the minimum spanning tree (MST) \cite{Wu}.
The importance of such a tree is that it identifies the
substructure of the network that is dominant for transport.
Starting from a completely empty network we insert links in decreasing
order of current magnitude, provided that they do not form a loop.
The resulting MST tree for the PIN is presented 
in Fig.~\ref{FIG_mst}, where the thickness of the links in the drawing
increases logarithmically with increasing current through a link.
The color of a node corresponds to the function performed by a protein
in the network. This tree (created solely on the base of current
flowing through a network) exhibits a large degree of modularity
where nodes that perform similar functions are close to each other.
It is also possible through this construction to identify the most critical links in the
network in terms of the largest current flowing through them.

Thus, the emerging picture from the above analysis for the PIN is one of a
network with a strong backbone that carries most part of the flow
combined with loops organized mainly within modules, so that flow through
this backbone is not really influenced. This result highlights the strong
modularity in the PIN structure. If a structure has a smaller degree of internal organization,
as is the case in the metabolic network, then flow is more uniformly distributed.

\section{Summary}
Summarizing, we have presented scaling arguments and simulations on
a class of self-similar complex networks, concerning transport on these
networks. Diffusion and resistance in these biological networks is important, since
they are both estimates of how many paths connect two nodes and how long these paths are.
By using the self-similarity property and a network renormalization scheme
we have developed a scaling theory of the resistance and diffusion dependence
on both the distance between two nodes and their corresponding degree.
We were able to recover a relation between network modularity and transport,
while the flow distribution in these networks was found to be
consistent with earlier studies, using different approaches.

\section{Methods}

\subsection{Resistance measurements}
In order to measure the conductivity between two nodes A and B we consider that all links in the
underlying network between any two neighbor nodes i and j have unit resistances $R_{ij}=1$.
By fixing the input current to $I_A=-1$ and the output to $I_B=1$
we can solve the Kirchhoff equations and compute the voltages in the system.
The measured resistance is then $R_{A\to B}=V_A-V_B$. However, due to the required inversion
of the relevant matrices we are limited by the computer resources to networks of relatively
small size, i.e. $N<10^4$ nodes.

In principle, the magnitude of $I_{ij}$ depends on the selection of the current
input/output nodes. Upon closer inspection, though, we found that the distribution of the current magnitudes
in the network links is not very sensitive to the selection of the current source and the
current sink. Comparison of the result of averaging over one input and many outputs and
over twenty different input and output pairs of nodes for the metabolic network showed that within
statistical error these two results are almost identical.

\subsection{Diffusion measurements}
In many cases (and especially those including real-life networks)
direct measurements of diffusion on complex networks exhibiting the small-world property may
present significant difficulties, due to the limited time-range where diffusion takes place
before settling quickly to a distance equal to the typical (very short) network diameter.
The rising part of the mean-squared displacement as a function of the time is very small
and reliable measurement of the diffusion exponent is very hard to do.
Moreover, we need additional information in order to quantify the $k$ and $\ell$
dependence of the diffusion time between two nodes.
For this purpose we used the peak of the first passage time distribution
as a typical diffusion time $T(A,B)$ between two points A and B in the network. Since this quantity may
be assymetric depending on which node we consider as origin, our diffusion time $T$
represents the average of $(T_{A\to B}+T_{B\to A})/2$.

\begin{acknowledgments}
The authors acknowledge support from NSF grants.
\end{acknowledgments}



\end{document}